\documentclass[aps,prl,reprint,superscriptaddress]{revtex4-2}

\usepackage{amsmath,amssymb}
\usepackage{graphicx} 
\usepackage{dcolumn} 
\usepackage{bm}       
\usepackage{hyperref}
\usepackage{xcolor}

\def\beq{\begin{equation}}
\def\eeq{\end{equation}}
\def\bc{\begin{cases}}
\def\ec{\end{cases}}
\def\bal{\begin{aligned}}
\def\eal{\end{aligned}}

\begin{document}

\title{Wormhole fragmentation and the axion Weak Gravity Conjecture}

\author{Puxin Lin}
\email{plin73@wisc.edu}

\author{Gary Shiu}
\email{shiu@physics.wisc.edu}
\affiliation{Department of Physics, University of Wisconsin-Madison, Madison, WI 53706, USA}

\begin{abstract}
We derive the axion Weak Gravity Conjecture (aWGC) from the nonperturbative fragmentation of Euclidean wormholes into charged defects. Requiring fragmentation to lower the Euclidean action fixes the axion action-to-charge bound and its order-one coefficient. Without additional scalars, this coefficient coincides with that of the Imaginary Distance Bound (IDB), while additional scalars strengthen the bound. The aWGC in the presence of multiple axio-dilaton pairs is discussed, leading to geometric representations. Axions satisfying the aWGC render the corresponding wormholes unstable, potentially suppressing their proliferation and resolving associated factorization problems.
\end{abstract}

\maketitle

Axions are ubiquitous in particle physics and quantum gravity and have rich phenomenological and cosmological implications, ranging from the strong CP problem and dark matter to inflation and astrophysics \cite{Hebecker:2018ofv,Marsh:2015xka}. A central question is therefore whether quantum gravity imposes universal constraints on the axion parameter space. The Weak Gravity Conjecture (WGC) \cite{Arkani-Hamed:2006emk}, originally formulated for $1$-form gauge fields, provides a natural framework for such constraints. An axionic Weak Gravity Conjecture (aWGC) can be motivated through duality transformations or dimensional reduction \cite{Brown:2015iha,Brown:2015lia,Heidenreich:2015nta}, or by considering gravitational instantons \cite{Montero:2015ofa,Hebecker:2016dsw}. Which class of gravitational instantons—cored, extremal, or Euclidean wormholes—underlies the aWGC has remained an open question \cite{Harlow:2022ich}, despite evidence favoring Euclidean wormholes \cite{Andriolo:2020lul,Andriolo:2022rxc}. Resolving this issue determines not only the origin and precise form of the aWGC bound, with implications for phenomenology, but also whether it can be understood as a consistency condition on gravitational saddles. 

Euclidean wormholes arise as gravitational instantons in theories of axions coupled to gravity \cite{Giddings:1987cg}. Such axion-gravity systems arise naturally in string theory \cite{Giddings:1989bq,Arkani-Hamed:2007cpn}, with explicit ten-dimensional uplifts of the wormhole solutions constructed in \cite{Loges:2023ypl}. As saddles of the Euclidean gravitational path integral (GPI), wormholes can mediate topology change and the nucleation of baby universes \cite{Hawking:1988ae,Coleman:1988tj,Giddings:1988wv}, raising the fundamental question of what their role is in the GPI \cite{Hebecker:2018ofv}.

The validity of wormholes as Euclidean saddles has consequently been scrutinized through analyses of their stability. Although potential negative modes were previously identified \cite{Rubakov:1996cn,Kim:2003js,Hertog:2018kbz}, recent work has established the perturbative stability of the wormhole solutions \cite{Loges:2022nuw,Jonas:2023qle,Hertog:2024nys,Marolf:2025evo}. Their persistence as Euclidean saddles sharpens the associated factorization problem \cite{Maldacena:2004rf}. Various approaches have been proposed, including ensemble interpretations \cite{Saad:2019lba,Cotler:2020ugk,Marolf:2020xie}, additional gravitational saddles \cite{Saad:2021rcu}, and Lorentzian formulations of the GPI \cite{Held:2026huj}, but the underlying issue remains open.

Wormholes pose a second challenge: even when individually well-defined as saddles, they can proliferate in the GPI. This has motivated the recently proposed Imaginary Distance Bound (IDB) \cite{Maldacena:2026jqd,DiUbaldo:2026rly}. From the gravitational-path-integral perspective, the IDB is associated with the absence of problematic wormholes in the low-energy theories \cite{Maldacena:2026jqd}; from an amplitude perspective, it has been argued that positivity is restored when wormholes become nonperturbatively unstable \cite{DiUbaldo:2026rly}. These observations suggest a natural question: can the instability of wormholes itself impose a known quantum-gravity bound on axions?

In this Letter, we show that the answer is affirmative. We find that axion wormholes can undergo nonperturbative fragmentation into charged zero-dimensional defects. Requiring the fragmentation channel to lower the Euclidean action yields the aWGC, including its order-one coefficient.
The aWGC thus emerges directly as a condition for the nonperturbative stability of gravitational saddles, without relying on indirect arguments based on duality symmetries or a higher-dimensional origin. The resulting bound fixes the order-one coefficient in the aWGC in terms of the action and charge of a probe defect, in a regime where the nonperturbative instability is under control.

We further establish a direct connection between this fragmentation bound and the IDB. In the absence of additional scalar fields, the two bounds coincide, while in the presence of additional scalars the fragmentation criterion is stronger, yielding a lower bound on the allowed action-to-charge ratio. Thus, whenever the aWGC is satisfied, the corresponding wormhole saddle is nonperturbatively unstable to fragmentation. This provides a possible mechanism for removing wormhole contributions that would otherwise lead to proliferation and factorization problems in GPIs.

\section{Formulation of the aWGC}

The WGC is a statement of stability of gravitational bound states, and for extremal black holes, it requires the existence of a decay channel. It was shown from the dynamics of charged fields in \cite{Lin:2024jug} that a non-zero decay rate leads to the constraint on the charge-to-mass ratio of the charged particle.

When the stability statement is applied to gravitational instantons instead of black holes, one expects the aWGC to follow. We show in the following that defects carrying axion charges are responsible for the fragmentation of wormholes. 
The wormhole-defect system accounts for the crucial gravitational effects in WGC-type arguments which appear to be absent between two probe instantons considered in 
\cite{Hattab:2026ief}.

Consider solutions to the following $D$-dimensional action in the axion picture 
\beq \label{eq:S_axion}
S=\int d^Dx\sqrt{g}\left[\frac{M_P^{D-2}}{2}\left(-R+\partial\phi^2\right)-\frac{f^2e^{\beta\phi}}{2}\partial \theta^2\right].
\eeq
We will set $M_P^{D-2}=1$ in this Letter and only recover it when necessary. When $\beta<\beta_c\equiv 2\frac{\sqrt{D-2}}{\sqrt{D-1}}$, wormhole solutions are admitted
\beq \bc
ds^2=\frac{dr^2}{1-\left(\frac{a_0}{r}\right)^{2(D-2)}}+r^2d\Omega_{D-1}^2\\
e^{\beta\phi}=\frac{Q^2}{(D-1)(D-2)a_0^{2(D-2)}f^2}\cos^2\left[\frac{\beta}{\beta_c}\arccos{\left(\frac{a_0}{r}\right)^{D-2}}\right]\\
\theta=\frac{(D-1)a_0^{D-2}}{Q}\frac{\beta_c}{\beta}\tan{\left[\frac{\beta}{\beta_c}\arccos{\left(\frac{a_0}{r}\right)^{D-2}}\right]}\\
\ec .\eeq
The wormhole has a throat located at $r=a_0$ and asymptotes to Euclidean space at $r\rightarrow \infty$. We normalize the dilaton and axion to zero at the wormhole throat, making $f$ the decay constant measured there and fixing the throat size to be $a_0^{2(D-2)}=\frac{Q^2}{(D-1)(D-2)f^2}$. Along the wormhole, the dilaton and axion follow a timelike geodesic in the moduli space,
\beq \label{eq:geodesic}
e^{-\beta\phi}-\left(\frac{\beta f}{2}\theta\right)^2=1.
\eeq

To see the fragmentation of the wormhole, we insert a defect in the background spacetime. A probe (Euclidean) $p$-brane coupled to a $(p+1)$-form field strength has an action of the form
\beq \bal
S&=\int d^{p+1}\sigma \mathcal{T}(\phi) \sqrt{-\det{\gamma_{ab}}}\\
-&\frac{iq}{(p+1)!}\int d^{p+1}\sigma \epsilon^{a_0 \cdots a_p}\partial_{a_0}X^{\mu_0}\cdots \partial_{a_p}X^{\mu_p}A_{\mu_0\cdots \mu_p},
\eal \eeq
where $\gamma_{ab}$ is the induced metric on the world-volume. From this, we write down the action of a $(-1)$-brane Euclidean axion defect as 
\beq \bal
S(x)=\mathcal{T}[\phi(x)]-q\theta(x)\\
=S_0 e^{-\frac{\beta}{2}\phi(x)}-q\theta(x),
\eal \eeq
where the absence of $i$ ensures consistency with the convention in (\ref{eq:S_axion}).
In the wormhole background, we can use the moduli space geodesic to write
\beq
\frac{\beta f }{2}\theta=\sqrt{e^{-\beta\phi}-1},
\eeq
and the defect action takes the form
\beq
S(x)=S_0e^{-\frac{\beta}{2}\phi}-\frac{2q}{\beta f }\sqrt{e^{-\beta\phi}-1}.
\eeq
The equation of motion sets the defect location such that the probe action is minimized. If this minimized value is negative, the wormhole action is lowered by the insertion of a defect and the wormhole is unstable. Since fixed charge wormholes have been shown to be perturbative stable \cite{Loges:2022nuw,Jonas:2023qle,Hertog:2024nys,Marolf:2025evo}, the instability due to defects is non-perturbative - its action is a local but not global minimum.

We proceed to associate the onset of such non-perturbative instability of wormholes, or better referred to as wormhole fragmentation, to the aWGC bound. To identify the onset, we assume there exists a point at which the action is negative. We show that a negative minimum always occurs at $r\rightarrow\infty$. Varying the action with respect to the location $x$, or the radial distance $r$, one finds
\beq \bal
S'(r)&=-\frac{\beta}{2}\phi'S_0e^{-\frac{\beta}{2}\phi}-\frac{2q}{\beta f }\frac{1}{2}\frac{-\beta\phi'}{\sqrt{e^{-\beta\phi}-1}}\\
&=c_1\left[S(r)-c_2\right],
\eal \eeq
where
\beq \bc
c_1=-\frac{\beta}{2}\phi'\\
c_2=\frac{2q}{\beta f}\frac{1}{\sqrt{e^{-\beta\phi}-1}}
\ec. \eeq
One can easily check that $c_{1,2}>0$ on $r\in(a_0,\infty)$, which implies that if $S(r)$ reaches zero, it monotonically decreases afterwards, such that a negative minimum can only occur at the asymptotic region. Applying $e^{\beta\phi_\infty}=\cos^2{\left(\frac{\beta}{\beta_c}\frac{\pi}{2}\right)}$, the asymptotic value of the defect action is found to be
\beq
S_*=\frac{S_0}{\cos{\left(\frac{\beta}{\beta_c}\frac{\pi}{2}\right)}}-\frac{q}{f}\frac{2}{\beta_c}\frac{\beta_c}{\beta}\tan{\left(\frac{\beta}{\beta_c}\frac{\pi}{2}\right)}.
\eeq
Requiring $S_*<0$ means that the wormhole action is lowered by insertion of the defect and that the wormhole is subject to fragmentation into smaller and eventually microscopic pieces. This will provide the aWGC bound on the axion action-to-charge ratio, which reads
\beq
\frac{f^\text{phy} S_0^\text{phy}}{q}<c_\text{WH}(\beta) M_P^\frac{D-2}{2},
\label{eq:aWGC}
\eeq
where $f^\text{phy}\equiv fe^\frac{\beta\phi_\infty}{2}, S_0^\text{phy}\equiv S_0e^{-\frac{\beta\phi_\infty}{2}}$ are the physical axion decay constant and tension at the defect location and
\beq
c_\text{WH}(\beta)=\frac{2}{\beta_c}\frac{\beta_c}{\beta}\sin{\left(\frac{\beta}{\beta_c}\frac{\pi}{2}\right)}.
\eeq
A probe D-instanton as considered in \cite{VanRiet:2020pcn}
satisfies, although not saturating (\ref{eq:aWGC}), allowing it to fragment the wormhole.

\subsection{Imaginary distance bound}
The form of the aWGC can be connected with geometric properties of the moduli space. We show that the form of constant factor $c_\text{WH}$ appearing in the aWGC can be rather universally expressed in terms of the moduli space distance $\Delta\tau$, which clarifies the relation between the aWGC and IDB.

We first notice that using the defect action $S_*<0$, the constant factor can be expressed as a direction in the moduli space,
\beq
c_\text{WH}=\frac{f\theta_\infty}{e^{-\frac{\beta\phi_\infty}{2}}},
\eeq
where the subscript labels values at the asymptotic region of the wormhole.

We consider an AdS moduli space with metric
\beq
ds^2=d\phi^2-f^2e^{\beta\phi}d\theta^2.
\eeq
As before, the moduli are normalized to $(e^{-\frac{\beta\phi_0}{2}}, \theta_0)=(1,0)$ at the wormhole throat. For massless scalars, the moduli follow geodesics emanating from this point,
\beq
e^{-\beta\phi}-\left(\frac{\beta f \theta}{2}\right)^2=1.
\eeq
The path can be parametrized by $e^{\beta\phi}=\cos^2{\lambda}$ and $\frac{\beta f}{2}\theta=\tan{\lambda}$ with affine parameter $\lambda$ and a straightforward calculation of the time-like distance yields
\beq
\Delta\tau=\int i ds=\frac{2}{\beta}\int d\lambda=\frac{4}{\beta}\lambda_\infty.
\eeq
This allows us to express the moduli space directly with the geodesic distance to find
\beq \label{eq:beautiful}
c_\text{WH}=\frac{2}{\beta}\sin{\left(\frac{\beta}{2}\frac{\Delta\tau}{2}\right)}.
\eeq
The above expression is universal in the sense that it holds for any wormhole solution as long as the moduli space is AdS and the moduli have no potential (remain massless). Of course, specific wormhole solutions will specify a particular value for $\Delta\tau$, e.g., $\frac{\Delta\tau}{2}=\frac{\pi}{\beta_c}$ for asymptotically flat wormholes and $\frac{\Delta\tau}{2}=\frac{\pi}{2}\frac{\sqrt{D-2}}{\sqrt{D-1}}$ for AdS wormholes, consistent with \cite{Maldacena:2026jqd, DiUbaldo:2026rly} upon restoring $M_P^{D-2}=(8\pi G_N)^{-1}$.

It is worthwhile to point out that the terms in (\ref{eq:beautiful}) carry geometrical meanings. $\frac{2}{\beta}$ is the radius of the moduli space and the argument inside the sine function is the ratio between the time-like distance and the moduli space radius. One can understand $c_\text{WH}$ as an effective distance of the wormhole when the moduli space is negatively curved. In the limit $\beta\rightarrow 0$, the dilaton decouples and the moduli space becomes flat and we find $c_\text{WH}=\frac{\Delta \tau }{2}$. More generally, on the negatively curved AdS moduli space, $c_\text{WH}\le\frac{\Delta \tau }{2}$. The aWGC being a stricter bound than the IDB has also been observed in some string theory constructions \cite{Etheredge:2026rio}. 

\section{Generalizations \label{sec:generalizations}}
\subsection{Extremal instanton}
Wormhole solutions cease to exist when $\beta>\beta_c$. There the gravitational solutions are cored instantons with curvature singularity and extremal instantons with a flat metric and divergent scalar field profile, which both exist for all values of $\beta$. Cored instantons require resolutions from UV physics near its core, and therefore we do not discuss them in this Letter. On the other hand, despite the divergent scalar field, extremal instanton has smooth metric and finite stress-energy tensor and can be used to derive an aWGC bound similar to the case of wormholes.

The moduli follow null geodesics in the case extremal instantons. The path can be parametrized by
\beq \bc
e^{\beta\phi}=\frac{A}{(1+\lambda)^2}\\
\theta=\frac{2}{\beta f}\frac{\lambda+1}{A}
\ec, \eeq
where $A\propto \frac{\beta Q}{f}$ and $\lambda=-1,0$ corresponds
to the center of the extremal instanton and asymptotic region respectively. Similar to the case of wormholes, a negative defect action is achieved at the asymptotic region if $S_0<\frac{q}{f}\frac{2}{\beta}$. We identify $c_\text{ext}=\frac{2}{\beta}$.

When wormholes exist, $\beta\le\beta_c$ and $c_\text{WH}<c_\text{ext}$, the aWGC is fixed by wormholes; when wormholes are forbidden, $\beta>\beta_c$, the aWGC bound is set by extremal instantons.

\subsection{Multiple Axio-dilaton pairs}

Axio-dilaton pairs commonly occur in the context of compactification. The aWGC has a natural generalization to cases with multiple decoupled pairs of axio-dilatons $(\phi_i,\theta_i)$.

Consider a wormhole carrying a vector of axion charges $\vec{Q}=(Q_1,\cdots,Q_n)$. The geodesic in the total product moduli space is the product of geodesics in individual moduli spaces,
\beq \bc
e^{\beta_i\phi_i}=\cos^2{( v_i\tau)}\\
\theta_i=\frac{2}{\beta_if_i}\tan{(v_i\tau)},
\ec \eeq
where $\tau$ is the timelike distance along the curve and $v_i$ is the projected speed satisfying
\beq \bc
\sum_i \frac{v_i^2}{\beta_i^2}=\frac{1}{4}\\
\frac{v_if_i}{\beta_iQ_i}=\text{const}\\
\ec. \eeq
The traversed moduli space distance determined by the wormhole geometry must be smaller than the moduli space diameter, requiring 
\beq
\frac{\Delta\tau}{2}<\frac{2}{\beta_\text{eff}}\equiv\sqrt{\sum \frac{4}{\beta_i^2}}.
\eeq

The probe defect action is extended to
\beq
S=S_0e^{-\frac{1}{2}(\sum \beta_i\phi_i)}-\sum q_i\theta_i,
\eeq
where now an important subtlety arises. For a single axio-dilaton pair, the combination $f^\text{phy}S_0^\text{phy}$ does not change with location. For a wormhole carrying multiple axion charges $\vec{Q}$, because $S_0^\text{phy}$ now couples to all dilatons, $f_i^\text{phy}S_0^\text{phy}$ runs when evaluated at different locations. It is therefore important that when imposing the aWGC, we use the physical parameters at the defect location.

Following the wormhole fragmentation argument, we find for multiple axion charges case the aWGC bound
\beq
S_0^\text{phy}<\sum \frac{q_i}{f_i^\text{phy}}\frac{2}{\beta_i}\sin{\left(\frac{\beta_i}{2}\frac{\Delta\tau}{2M_P^\frac{D-2}{2}}\Delta_i\right)}.
\eeq
Here $\Delta_i=\frac{f_i^{-1}Q_i}{\sqrt{\sum f_i^{-2}Q_i^2}}$ describes the partition of the total moduli distance into the individual $AdS_2$ moduli subspace. $\Delta_i$ should satisfy $\frac{\beta_i}{2}\frac{\Delta\tau}{2}\Delta_i\le \frac{\pi}{2}$ so that the partitioned moduli space distance can fit in the subspace. We can further define the charge-to-action ratio $z_i\equiv\frac{q_i}{f_i^\text{phy}S_0^\text{phy}}$ and the wormhole $Y$-vector $Y_i=\frac{2}{\beta_i}\sin{\left(\frac{\beta_i}{2}\frac{\Delta\tau}{2}\Delta_i\right)}$ so that the aWGC is compactly expressed as
\beq \label{eq:dot_product}
\vec{z}\cdot\vec{Y}>1. 
\eeq
The above expression is reminiscent of the particle WGC with multiple U(1) charges formulated with the black hole decay or black hole-particle repulsion argument \cite{Lin:2025wfe,Lin:2025gco}.

The condition to fragment an extremal instanton is analogously computed and takes exactly the form of (\ref{eq:dot_product}). The $Y$-vectors associated to extremal instantons are $\vec{Y}=(\frac{2b_1}{\beta_1},\cdots,\frac{2b_n}{\beta_n})$ with $b_i=0,1$ can be individually turned on and off.

We now return to the formulation of the aWGC for cases with multiple axion charges. The aWGC requires existence of a set of charged axion defects $\{\vec{z}_I\}$ such that every $\vec{Y}(\vec{Q})$ satisfies (\ref{eq:dot_product}) for some $\vec{z}_I$. We will provide two geometric pictures of the statement, 1) an equivalent condition phrased with polar sets and 2) a sufficient condition using convex hull.

\begin{figure}
    \centering
    \includegraphics[width=0.5\linewidth]{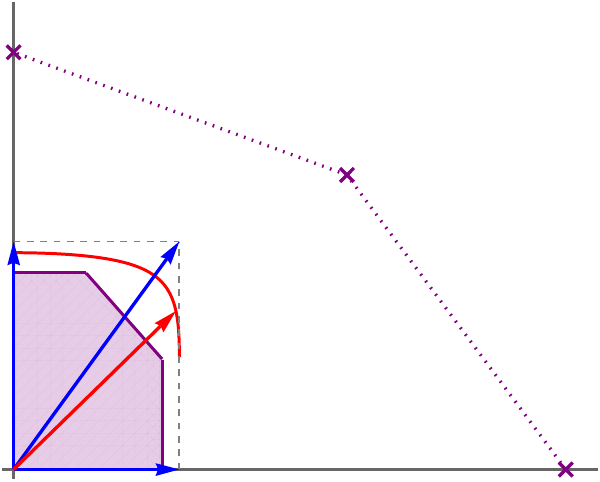}
    \caption{Polar set condition. The purple crosses represent the axion defects $\{\vec{z}_I\}$ and the purple shade corresponds to the polar set $Z^\circ$. The blue arrows and red solid curve correspond, respectively, to the $Y$-vectors of the extremal instantons and wormholes with all possible charge vector $\vec{Q}$. The aWGC is satisfied if and only if all the $Y$-vectors lay outside $Z^\circ$.}
    \label{fig:PolarSet}
\end{figure}

\textbf{Polar set condition}: Given the collection of axions in a theory, $Z\equiv\{\vec{z}_I\}$, its polar set is defined as $Z^\circ\equiv \{\vec{x}|\vec{x}\cdot\vec{z}\le 1, \forall \vec{z}\in Z\}$. One can interpret the polar set as objects that do not fragment into axion defects. The aWGC is therefore equivalent to requiring that no wormhole or extremal instanton has a $Y$-vector residing in the polar set, i.e.,
$\{\vec{Y}\}\subset \overline{Z^\circ}$. The geometry is given in Figure. \ref{fig:PolarSet}.

\begin{figure}
    \centering
    \includegraphics[width=0.5\linewidth]{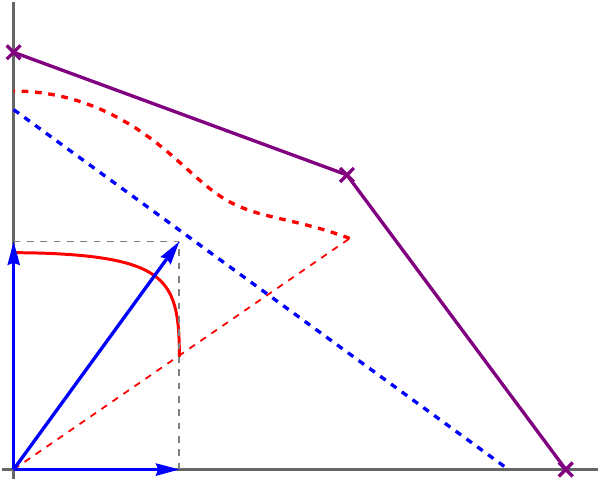}
    \caption{Convex hull condition. The red and blue dotted regions correspond to the reciprocal $\tilde{Y}$ of the $Y$-vectors of wormholes and extremal instantons respectively. The purple boundary is the convex hull spanned by $\{\vec{z}_I\}$ and it covers $\tilde{Y}$, satisfying the aWGC. Note that only the first quadrant is represented here, but $\{\vec{z}_I\}$ contains all anti-particles as well.}
    \label{fig:ConvexHull}
\end{figure}

\textbf{Convex hull condition}:
From the $Y$-vectors of wormholes and extremal instantons, define the reciprocal set, $\tilde{Y}\equiv\{\vec{x}|\vec{x}=\frac{\vec{y}}{|\vec{y}|^2},\vec{y}\in \{\vec{Y}\}\}$. The aWGC will be satisfied if $\tilde{Y}$ resides completely inside the convex hull spanned by $\{\vec{z}_I\}$, as shown in Figure. \ref{fig:ConvexHull}.

In Figures \ref{fig:PolarSet} and \ref{fig:ConvexHull}, We provide the geometric representations of the aWGC for $n=2$ axio-dilaton pairs. To showcase all possible features, we choose $\beta_1>\beta_2$ satisfying $\frac{2}{\beta_1}<\frac{\Delta\tau}{2}<\frac{2}{\beta_2}<\frac{2}{\beta_\text{eff}}$. The first inequality implies that the moduli subspace of the first axio-dilaton pair $(\phi_1,\theta_1)$ alone cannot fit the wormhole. Therefore, the red curve corresponding to the $Y$-vector of all possible wormholes does not cover angles near the $\theta_1$ direction. When the wormhole in some charge direction does not exist, one would need to resort to the extremal instantons to identify the aWGC condition and when wormholes exist, they always lead to a stricter condition, as can be seen in both Figures \ref{fig:PolarSet} and \ref{fig:ConvexHull}.

We further discuss the relation of the IDB to the aWGC with multiple axion charges. To find the analogous quantity $c_\text{WH}$ in this case, we note that a necessary condition for the aWGC to be satisfied is that there must exist $\vec{z}_I$ such that $|\vec{z}_I|\cdot\min{\{|\vec{Y}|\}}>1$. We can therefore define, for multiple axio-dilaton pairs, $c_\text{WH}\equiv \min{\{|\vec{Y}|\}}$. For wormholes, $\sum \Delta_i^2=1$, and
\beq
|\vec{Y}|^2=\sum \frac{4}{\beta_i^2}\sin^2{\left(\frac{\beta_i}{2}\frac{\Delta\tau}{2}\Delta_i\right)}\le\left(\frac{\Delta\tau}{2}\right)^2.
\eeq
This indicates that $c_\text{WH}\le\frac{\Delta\tau}{2}$ holds true when generalizing to multiple axion charges, and equality is achieved only when all $\beta_i\rightarrow 0$.

A major obstacle to extracting precise phenomenological implications of the aWGC has been the ambiguity in its order-one coefficient. Our result determines this coefficient by requiring wormhole fragmentation to lower the Euclidean action, in a regime where both the wormhole solution and the fragmentation analysis are under control. This provides a quantitative basis for sharpening quantum gravity constraints on axion models, with potential implications for a broad range of axion phenomenology and cosmology, including the strong CP problem, dark matter, inflation, and dark energy.

\begin{acknowledgments}
\textbf{Acknowledgments} We thank Thomas Van Riet, Ben Freivogel, Jan de Boer, Jan Pieter van der Schaar and David Berenstein for useful feedback. We thank the hospitality of the IoP at the University of Amsterdam, at which part of this study was conducted. This work is supported in part by the DOE Award
DE-SC0017647, and a WARF Named Professorship, provided by the University of Wisconsin–Madison Office of the Vice Chancellor
for Research with funding from the Wisconsin Alumni Research Foundation. 
\end{acknowledgments}

\bibliography{references}

\end{document}